\documentclass{ws-mplb}
    \usepackage{bm}
    \usepackage{graphicx}

\newcommand\beq{\begin{equation}}
\newcommand\eeq{\end{equation}}
\newcommand\beqa{\begin{eqnarray}}
\newcommand\eeqa{\end{eqnarray}}
\newcommand{\nn}{\nonumber\\}

\begin{document}
\markboth{Andr\'es Santos}{Comment on the solution of the Percus--Yevick integral equation for square-well fluids}

\title{Comments on ``State equation for the three-dimensional system of `collapsing' hard spheres''}

\author{Andr\'es Santos}

\address{Departamento de F\'{\i}sica, Universidad de
Extremadura, E-06071 Badajoz, Spain\\
andres@unex.es}


\maketitle

\begin{abstract}
A recent paper [I. Klebanov et al.\ \emph{Mod. Phys. Lett. B} \textbf{22} (2008) 3153; arXiv:0712.0433]
claims that  the exact solution of the Percus--Yevick (PY) integral equation for a system of hard spheres plus a step potential  is obtained. The aim of this paper is to show that Klebanov et al.'s result is incompatible with the PY equation since it violates two known cases: the low-density limit and the hard-sphere limit.
\end{abstract}


\section{Introduction}
Given a fluid of particles interacting via a certain potential $\phi(r)$, access to its (equilibrium) structural and thermodynamic properties is usually obtained by means of approximate integral equations,\cite{HM06} whose solution typically requires hard numerical work. Exceptions are practically restricted to the Percus--Yevick (PY) equation for hard spheres\cite{W63,T63} and sticky hard spheres,\cite{B68} and the mean spherical approximation (MSA) for the hard-core Yukawa potential.\cite{W73}

The simplest potential including an energy scale and two length scales is that of hard spheres plus a step-function tail:
\beq
\phi(r)=
\begin{cases}
\infty,&r<\sigma,\\
\epsilon,&\sigma<r<\lambda,\\
0,&r>\lambda.
\end{cases}
\label{0}
\eeq
If $\epsilon<0$, the step potential is attractive and Eq.\ \eqref{0} describes the well-known square-well potential. On the other hand, $\epsilon>0$ defines the square-shoulder potential. The pure hard-sphere fluid is recovered if either $\epsilon=0$ (at arbitrary $\lambda/\sigma$), or $\lambda=\sigma$ (at arbitrary, but finite, $\epsilon$), or $\epsilon\to\infty$ (again, at arbitrary $\lambda/\sigma$).

Starting from the PY integral equation, Wertheim\cite{W64}  was able to express the Laplace transform
\beq
G(t)\equiv \int_\sigma^\infty dr\, e^{-tr} r g(r),
\label{6}
\eeq
where $g(r)$ is the radial distribution function, in terms of quantities involving the cavity function $y(r)\equiv e^{\phi(r)/k_BT}g(r)$ (where $k_B$ is the Boltzmann constant and $T$ is the temperature) only in the interval $0\leq r\leq \lambda$:
\beq
G(t)=\frac{(1+4\pi\rho K)t^{-2}-F(t)+2\pi\rho t^{-1}\left[Y(-t)-Y(t)\right]}{1+2\pi\rho t^{-1}\left[F(-t)-F(t)\right]},
\label{7}
\eeq
where $\rho$ is the number density, $K\equiv -F'(0)$, and
\beq
F(t)\equiv-\int_0^\lambda dr\, e^{-tr}r f(r) y(r),
\label{8}
\eeq
\beq
Y(t)\equiv-\int_0^{\lambda-\sigma}dr\, e^{-tr}\int_{\sigma+r}^\lambda dr'\, r'f(r')y(r')(r-r')\left[1+f(r-r')\right]y(r-r').
\label{9}
\eeq
Here, $f(r)\equiv e^{-\phi(r)/k_BT}-1$ is the Mayer function. Equations \eqref{7}--\eqref{9} apply not only to the potential \eqref{0} but more in general to any interaction with a hard-core at $r=\sigma$ and a finite range at $r=\lambda$.

In a recent paper,\cite{KGG08} Klebanov et al.\ claim that they obtain the exact solution of the PY integral equation for the potential \eqref{0}. According to their approach, Eq.\ \eqref{7} is complemented by
\beq
y(r)=C_1 +C_2 r+C_4 r^3,\quad 0\leq r\leq \lambda,
\label{5}
\eeq
where the coefficients $C_1$, $C_2$, and $C_4$ are the solutions of a closed set of equations. Inserting Eq.\ \eqref{5} into Eqs.\ \eqref{8} and \eqref{9}, one gets $F(t)$ and $Y(t)$, and hence $G(t)$ through Eq.\ \eqref{7}.

The aim of this paper is to show that, in contrast to what is claimed in Ref.~\cite{KGG08}, Eq.\ \eqref{5} is not compatible with the PY solution because it contradicts known results in the low-density limit as well as in the hard-sphere limit.

\section{Low-density limit}
The virial expansion of the cavity function is
\beq
y(r)=1+\sum_{n=1}^\infty \rho^n y_n(r),
\label{1}
\eeq
where the functions $y_n(r)$ are represented by sums of diagrams.\cite{HM06} In the special case of the potential \eqref{0}, the first-order contribution $y_1(r)$ is given by\cite{BH67}
\beq
y_1(r)=(1+\gamma)^2 \Phi_{\sigma,\sigma}(r)-2\gamma(1+\gamma)\Phi_{\sigma,\lambda}(r)+\gamma^2\Phi_{\lambda,\lambda}(r),
\label{2}
\eeq
where $\gamma\equiv e^{-\epsilon/k_BT}-1$ and
\beqa
\Phi_{a,b}(r)&\equiv&\frac{\pi}{12r}\left[3(a+b)^2-2(b-a)r-r^2\right](b-a-r)^2\Theta(b-a-r)\nn
&&-\frac{\pi}{12r}\left[3(b-a)^2-2(a+b)r-r^2\right](a+b-r)^2\Theta(a+b-r),
\label{3}
\eeqa
$\Theta(x)$ being Heaviside's step function. More explicitly, in the interval $0\leq r\leq 2\sigma$ one has
\beqa
y_1(r)&=&(1+\gamma)^2\frac{\pi}{12}(4\sigma+r)(2\sigma-r)^2
+\gamma^2\frac{\pi}{12}(4\lambda+r)(2\lambda-r)^2-2\gamma(1+\gamma)\frac{4\pi}{3}\sigma^3\nn
&&+
2\gamma(1+\gamma)\times\begin{cases}
0,&0\leq r\leq \lambda-\sigma,\\
\pi\frac{3(\sigma+\lambda)^2-2(\lambda-\sigma)r-r^2}{12r}(\lambda-\sigma-r)^2,&\lambda-\sigma\leq r\leq 2\sigma.
\end{cases}
\label{4}
\eeqa
We see that, while $y_1(r)$ is a cubic function in the interval $0\leq r\leq \lambda-\sigma$, it is a quartic polynomial function divided by $r$  for $ r> \lambda-\sigma$. Moreover, the second derivative of $y_1(r)$ is discontinuous at $r=\lambda-\sigma$. Therefore, Eq.\ \eqref{5} is inconsistent with Eq.\ \eqref{4} to first order in density. Since the PY theory yields the exact $y_1(r)$ for any interaction potential,\cite{HM06} we conclude that Eq.\ \eqref{5} is inconsistent with the true PY solution.

\section{Hard-sphere limit}
As an independent test, let us now take the limit $\epsilon\to 0$ keeping $\lambda/\sigma$ fixed. In that case, as said above, the potential \eqref{0} reduces to that of hard spheres of diameter $\sigma$, the value of $\lambda>\sigma$  not playing any role. The exact solution of the PY equation for hard spheres is well known.\cite{HM06,W63,T63,W64} In particular, the functional form of $y(r)$ for $0\leq r\leq 2\sigma$ is\cite{W63}
\beq
y(r)=\begin{cases}
A_0+A_1 r+A_3 r^3,&0\leq r\leq \sigma,\\
\frac{B_1}{r}e^{-\kappa_1 r}+\frac{B_2}{r}e^{-\kappa_2 r}\cos\left(\omega r+\varphi\right), & \sigma\leq r\leq 2\sigma,
\end{cases}
\label{10}
\eeq
where the coefficients $A_i$, $B_i$, $\kappa_i$, $\omega$, and $\varphi$ are functions of density whose explicit expressions will not be needed here.
It is quite clear that Eq.\ \eqref{5} cannot reduce to Eq.\ \eqref{10} if $\epsilon\to 0$ with $\lambda>\sigma$. Therefore, Eq.\ \eqref{5}   is again incompatible with the PY equation.

\section{Conclusion}
In summary, the approximation presented in Ref.~\cite{KGG08}  is not the solution of the PY integral equation for the interaction potential \eqref{0}, in contrast to what is claimed by  Klebanov et al. The flaw in their derivation could be due to the fact that, at a given point, the authors discard the product $G(t)F(-t)$ when manipulating Eq.\ \eqref{7}. Therefore, what they obtain is, at most, an approximation to the true solution of the PY equation, differing from the latter even in the low-density and in the hard-sphere limits.

\section*{Acknowledgments}
Partial support from the Ministerio de Educaci\'on y Ciencia (Spain)
through Grant No.\ FIS2007--60977 (partially financed by FEDER
funds) and from the Junta de Extremadura through Grant No.\ GRU09038
is acknowledged.

\end{document}